\documentclass[%
 preprint,
superscriptaddress,
amsmath,amssymb,
 aps, physrev,
]{revtex4-2}

\usepackage{graphicx}
\usepackage{dcolumn}
\usepackage[T1]{fontenc}
\usepackage[utf8]{inputenc}
\usepackage{bm}
\usepackage{xcolor}
\usepackage[normalem]{ulem}
\usepackage{braket}

\usepackage[linktocpage=true,
  colorlinks=true, 
  pdfborder={0 0 0},
  linkcolor=blue,
  citecolor=red,
  filecolor=yellow,
  urlcolor=blue,
  bookmarks,
  pdfauthor={},
]{hyperref}

\begin{document}

\title{Iterative learning scheme for crystal structure prediction with anharmonic lattice dynamics}

\author{Hao Gao} \affiliation{Centro de Física de Materiales (CFM-MPC), CSIC-UPV/EHU, Manuel de Lardizabal pasealekua 5, 20018 Donostia/San Sebastián, Spain}
\affiliation{Fisika Aplikatua Saila, Gipuzkoako Ingeniaritza Eskola, University of the Basque Country (UPV/EHU), Europa Plaza 1, 20018 Donostia/San Sebastián, Spain}

\author{Yue-Wen Fang}  \email{yuewen.fang@ehu.eus}\affiliation{Centro de Física de Materiales (CFM-MPC), CSIC-UPV/EHU, Manuel de Lardizabal pasealekua 5, 20018 Donostia/San Sebastián, Spain}

\author{Ion Errea}  \email{ion.errea@ehu.eus}\affiliation{Centro de Física de Materiales (CFM-MPC), CSIC-UPV/EHU, Manuel de Lardizabal pasealekua 5, 20018 Donostia/San Sebastián, Spain}
\affiliation{Fisika Aplikatua Saila, Gipuzkoako Ingeniaritza Eskola, University of the Basque Country (UPV/EHU), Europa Plaza 1, 20018 Donostia/San Sebastián, Spain}
\affiliation{Donostia International Physics Center (DIPC), Manuel de Lardizabal Pasealekua 4, 20018 Donostia/San Sebastián, Spain}

\date{\today}

\begin{abstract}
First-principles based crystal structure prediction (CSP) methods have revealed an essential tool for the discovery of new materials.
However, in solids close to displacive phase transitions, which are common in ferroelectrics, thermoelectrics, charge-density wave systems, or superconducting hydrides, the ionic contribution to the free energy and lattice anharmonicity become essential, limiting the capacity of CSP techniques to determine the thermodynamical stability of competing phases. 
While variational methods like the stochastic self-consistent harmonic approximation (SSCHA) accurately account for anharmonic lattice dynamics \emph{ab initio}, their high computational cost makes them impractical for CSP. 
Machine-learning interatomic potentials offer accelerated sampling of the energy landscape compared to purely first-principles approaches, but their reliance on extensive training data and limited generalization restricts practical applications. Here, we propose an iterative learning framework combining evolutionary algorithms, atomic foundation models, and SSCHA to enable CSP with anharmonic lattice dynamics. Foundation models enable robust relaxations of random structures, drastically reducing required training data.  Applied to the highly anharmonic H$_3$S system, our framework achieves good agreement with the benchmarks based on density functional theory, accurately predicting phase stability and vibrational properties from 50 to 200 GPa. Importantly, we find that the statistical averaging in the SSCHA reduces the error in the free energy evaluation, avoiding the need for extremely high accuracy of machine-learning potentials. This approach bridges the gap between data efficiency and predictive power, establishing a practical pathway for CSP with anharmonic lattice dynamics.

\end{abstract}

\maketitle

\section{Introduction}
\label{sec:intro}

Crystal structure prediction (CSP) coupled with density functional theory (DFT) is a fundamental tool for discovery of novel materials and leads to breakthroughs across diverse domains including superhard materials, energetic materials and exotic chemical compounds ~\cite{zhangMaterialsDiscoveryHigh2017,oganovStructurePredictionDrives2019,NPJComputational2019-superhard-Zurek,ML-CSP-JMI2024,Liu2023-exotic-Helium-NatComm}. A noteworthy success of CSP is its application in discovering high-temperature superconducting hydrides~\cite{PRL2017-YHx-YanmingMA,PRL2019-Li2⁢MgH16-YanmingMA,duanPressureinducedMetallizationDense2014,liuPotentialHighTcSuperconducting2017,LaH10-LaH16-PRB2020-Kruglov,PRB2020-CSH-ZurekHemley,CommMater2024-LuNH-Fang,SciAdv2025-molecular-hydrides-ZhaoLIU,PRB2021-LaBH8-Boeri}. Inspired by the progress of CSP in hydrogen-based superconductors, some hydrides such as H$_3$S~\cite{duanPressureinducedMetallizationDense2014} and LaH$_{10}$~\cite{liuPotentialHighTcSuperconducting2017,LaH10-LaH16-PRB2020-Kruglov} with superconducting critical temperatures beyond 200 K have been experimentally synthesized and confirmed at extreme pressures~\cite{drozdovConventionalSuperconductivity2032015,drozdovSuperconductivity250Lanthanum2019,Hemley_LaH_PRL2019,Osmond2022Clean}.

Despite these successes, a key limitation of conventional CSP is its neglect of quantum and thermal lattice vibrational effects in the energetics. Standard CSP workflows rank candidate structures by enthalpy after relaxation on the Born–Oppenheimer (BO) potential energy surface (PES), ignoring the kinetic energy of the ions to reduce computational costs. However, it breaks down when ionic lattice dynamics, highly influenced by the anharmonic part of the PES, plays an important role in the thermodynamic stability, as it is the case in systems close to displacive phase transitions such as thermoelectric materials~\cite{Aseginolaza2019Phonon,dangicLatticeThermalConductivity2025}, ferroelectrics or quantum paraelectrics~\cite{Ribeiro2018Strong,Verdi2023Quantum,Libbi2025Nonequilibrium,2015-SB-ferroelectrics}, charge-density wave systems~\cite{Bianco2020Weak,Diego2021vanderWaals,alkortaSymmetrybrokenGroundState2025}, etc. In all of these cases, CSP methods miss phases stabilized by quantum and thermal ionic fluctuations, given that none of them correspond to a local minimum of the PES.

This limitation is particularly evident in superconducting hydrides, where light hydrogen atoms are subject to strong quantum fluctuations that generate significant anharmonic effects~\cite{erreaQuantumHydrogenbondSymmetrization2016,erreaQuantumCrystalStructure2020,CommMater2024-LuNH-Fang,dangicAmbientPressureHigh2024,dangicAmbientPressureHigh2024,erreaHighPressureHydrogenSulfide2015,Hou2021Strong,Errea2013First-Principles,Rousseau2010Giant,Belli2022Impact}. Such anharmonicity substantially modifies free energy surfaces, altering both thermodynamic and dynamical phase stability from the conclusions drawn exclusively from the PES obtained within the BO approximation. For both H$_3$S and LaH$_{10}$ systems, the experimentally identified cubic phases that yield the highest superconducting critical temperatures~\cite{drozdovConventionalSuperconductivity2032015,drozdovSuperconductivity250Lanthanum2019,sun_high-temperature_2021} were theoretically predicted to be dynamically unstable at the relevant experimental pressures within the standard harmonic approximation (HA), characterized by imaginary phonon modes and higher BO enthalpies than competing distorted phases. Crucially, incorporation of anharmonic effects eliminates these imaginary modes and stabilizes the cubic structures thermodynamically~\cite{erreaHighPressureHydrogenSulfide2015,erreaQuantumCrystalStructure2020}, thus making possible the existence of these cubic structures with high electron-phonon coupling structures at lower pressures and raising a critical question of whether anharmonic effects should be incorporated into CSP to discover high superconducting critical temperature ($T_c$) hydrides at lower pressures. However, efforts to include anharmonic lattice dynamics in CSP remain scarce. Despite Kruglov \textit{et al.}~\cite{kruglovCrystalStructurePrediction2023} proposed a hybrid approach combining evolutionary algorithms (EA), molecular dynamics (MD), thermodynamic integration technique, and machine-learning interatomic potentials (MLIPs), a generally practical method for CSP that accounts for anharmonic effects is still lacking.

MLIPs are known to accelerate both SSCHA~\cite{dangicLatticeThermalConductivity2025,Zhao2025-npj-comput-mater-XYB6C6,DONG-LaBeH8-MatTodayPhys2025,alkortaSymmetrybrokenGroundState2025} and CSP calculations~\cite{podryabinkinAcceleratingCrystalStructure2019, tongAcceleratingCALYPSOStructure2018, wangConcurrentLearningScheme2024, wangMAGUSMachineLearning2023, pickardEphemeralDataDerived2022, salzbrennerDevelopmentsFurtherApplications2023, bisboEfficientGlobalStructure2020, pitfieldAugmentationUniversalPotentials2025,ML-CSP-JMI2024}. The features required from the potentials by these methods are, however, different. SSCHA requires MLIPs with high accuracy in
computing atomic forces and energies of structures in a small configuration space, defined by the quantum and thermal fluctuations of a particular compound. On the other hand, CSP requires general, robust, and transferable
MLIPs covering most of the configuration space, given that the structures generated during CSP are highly diverse and specific-purpose MLIPs usually require massive training data to avoid failure in local optimizations for randomly generated structures. Recently, multiple atomic foundation models pretrained on large datasets have been proposed \cite{dengCHGNetPretrainedUniversal2023,batatiaFoundationModelAtomistic2025,yangMatterSimDeepLearning2024}. These models enable ``robust'' MD simulations and geometry optimizations of various systems and random structures without additional training \cite{dengCHGNetPretrainedUniversal2023,batatiaFoundationModelAtomistic2025,yangMatterSimDeepLearning2024}. Moreover, finetuning with a small amount of system-specific data further enhances their accuracy for target systems \cite{yangMatterSimDeepLearning2024,pitfieldAugmentationUniversalPotentials2025}, bridging the gap between data efficiency and predictive power.  

Here, we propose an iterative learning scheme that combines EA and atomic foundation models, followed by SSCHA relaxations, with the aim of implementing CSP with anharmonic lattice dynamics. The application of foundation models dramatically reduces the required training data size. Crucially, we reveal that the computation of thermodynamical properties by SSCHA does not require MLIPs with ultrahigh accuracy. Applied to the highly anharmonic H$_3$S system \cite{duanPressureinducedMetallizationDense2014,drozdovConventionalSuperconductivity2032015,liDissociationProductsStructures2016,erreaQuantumHydrogenbondSymmetrization2016}, our framework achieves excellent agreement with DFT benchmarks, indicating its potential for practical CSP with anharmonicity.  

\section{\label{sec:sscha} Anharmonicity within the stochastic self-consistent harmonic approximation}

Before describing the iterative learning scheme to develop MLIPs accurate enough to perform CSP within the SSCHA, we find it convenient to review the latter theory. The SSCHA~\cite{erreaAnharmonicFreeEnergies2014,monacelliStochasticSelfconsistentHarmonic2021,biancoSecondorderStructuralPhase2017,Errea2013First-Principles} is a variational method used to obtain the structural and vibrational properties of materials at any temperature while explicitly including quantum ionic effects. The approach minimizes a trial free energy $\mathcal{F}[\tilde{\rho}]$, defined as a functional of a trial density matrix $\tilde{\rho}$. According to the Gibbs–Bogoliubov variational principle, the minimum value of $\mathcal{F}[\tilde{\rho}]$ provides an upper bound to the true free energy $F$ of the system:
\begin{equation}
    \mathcal{F}[\tilde \rho] = \braket{K + V}_{\tilde\rho}  - TS[\tilde \rho] \ge   F.
    \label{eq:bogo}
\end{equation}
Here, $\braket{K + V}_{\tilde\rho}$ represents the quantum statistical average of the ionic kinetic energy $K$ and the BO energy $V$, while the second term accounts for the entropic contribution, with $S$ being the entropy associated to the ions and $T$ the temperature. The notation $\braket{O}_{\tilde\rho}$ denotes the quantum statistical average of an operator $O$ evaluated using the trial density matrix $\tilde{\rho}$.

The trial density matrix of the SSCHA is approximated as the one described by an auxiliary harmonic Hamiltonian:
\begin{equation}
    \mathcal{H}_{\bm{\mathcal{R}},\bm{\Phi}} = K + \frac 12 \sum_{ab}(R_a - \mathcal{R}_a) \Phi_{ab} (R_b - \mathcal{R}_b).
    \label{eq:HSCHAtrial}
\end{equation} 
Here $\bm{\mathcal{R}}$ are the centroid positions, which correspond to the average atomic positions, and $\bm{\Phi}$ the auxiliary force-constant matrix, which determine the amplitude of the fluctuations of the ions around $\bm{\mathcal{R}}$. $\bm{R}$ denotes all ionic positions in one single vector, and the indices $a$ and $b$ run over both atomic and Cartesian coordinates (from 1 to $3N_a$, with $N_a$ the number of atoms in the cell). The ionic probability distribution defined by the auxiliary harmonic Hamiltonian is a Gausian written as
\begin{equation}
    \tilde \rho_{\bm{\mathcal{R}},\bm{\Phi}}(\bm{R}) = \sqrt{\bm{\det(\bm{\Psi}}^{-1}/2\pi)} \times \exp \left[-\frac 12 \sum_{ab}(R_a - \mathcal{R}_a) \Psi_{ab} (R_b - \mathcal{R}_b) \right],
    \label{eq:gauss}
\end{equation}
where $\bm{\Psi}$ is the displacement-displacement correlation matrix
\begin{equation}
    \Psi_{ab} = \left \langle {u_au_b}\right \rangle_{\tilde \rho_{\bm{\mathcal{R}},\bm{\Phi}}},
    \label{eq:psi0}
\end{equation}
with displacements $\bm{u} = \bm{R} - \bm{\mathcal{R}}$. Given that the trial density matrix is fully parametrized by the centroid positions  $\bm{\mathcal{R}}$ and the auxiliary force constants $\bm{\Phi}$, the SSCHA minimizes the
\begin{equation}
\mathcal{F}[\bm{\mathcal{R}},\bm{\Phi}]=\braket{K + V}_{\tilde\rho_{\bm{\mathcal{R}},\bm{\Phi}}}  - TS[\tilde \rho_{\bm{\mathcal{R}},\bm{\Phi}}]
\label{eq:fvariational}
\end{equation}
functional with respect to these variables. At the end of the minimization, not only a good approximation of the free energy is obtained, the obtained centroid positions also correspond to the renormalized average positions of the ions considering thermal and quantum ionic fluctuations and anharmonicity. The SSCHA can also effectively relax lattice parameters at any pressure considering lattice anharmonicity~\cite{Monacelli2018Pressure}.

The SSCHA free energy minimization is implemented with a preconditioned gradient-descent algorithm~\cite{monacelliStochasticSelfconsistentHarmonic2021}, which requires the calculation of the 
\begin{equation}
\frac{\partial \mathcal{F}}{\partial \mathcal{R}_a}=-\left \langle \text{f}^{BO}_a(\bm{R})  \right \rangle_{\tilde \rho_{\bm{\mathcal{R}},\bm{\Phi}}}
\label{eq:gradR}
\end{equation}
and 
\begin{equation}
    \mathcal{G}^{\bm{\Phi}}_{ab} = \left \langle \left( \text{f}^{BO}_b(\bm{R}) - \text{f}^{\mathcal{H}_{\bm{\mathcal{R}},\bm{\Phi}}}_b(\bm{R})\right)\sum_e \Psi^{-1}_{ae} \left(R_e -  \mathcal{R}_e \right) \right \rangle_{\tilde \rho_{\bm{\mathcal{R}},\bm{\Phi}}}
    \label{eq:pregradphi}
\end{equation}
quantum statistical averages. Here $\textbf{f}^{BO}_b(\bm{R})$ are the Born-Oppenheimer forces on the ions, evaluated either from DFT or MLIP in this work, while $\textbf{f}^{\mathcal{H}_{\bm{\mathcal{R}},\bm{\Phi}}}_b(\bm{R})$ are the forces from the auxiliary harmonic Hamiltonian $\mathcal{H}_{\bm{\mathcal{R}},\bm{\Phi}}$. These averages are calculated stochastically as
\begin{equation}
\frac{\partial \mathcal{F}}{\partial \bm{\mathcal{R}}}= - \frac{1}{N_c} \sum_i^{N_c} \textbf{f}^{BO}(\bm{R}_i),
\label{eq:gradR_simple}
\end{equation}
and
\begin{equation}
    \mathcal{G}^{\bm{\Phi}}_{ab} = \frac{1}{N_c}\sum_i^{N_c} \mathcal{G}^{\bm{\Phi}}_{i,ab},
    \label{eq:pregradphi_average}
\end{equation}
where
\begin{equation}
    \mathcal{G}^{\bm{\Phi}}_{i,ab} =  \left( \text{f}^{BO}_b(\bm{R}_i) - \text{f}^{\mathcal{H}_{\bm{\mathcal{R}},\bm{\Phi}}}_b(\bm{R}_i)\right)\sum_e \Psi^{-1}_{ae} \left(R_{i,e} -  \mathcal{R}_e \right).
    \label{eq:pregradphi_single}
\end{equation}
The set of ionic configurations $\{\bm{R}_i\}$ are created according to the Gaussian distribution in Eq. \eqref{eq:gauss}. In Eqs. \eqref{eq:pregradphi_average} and \eqref{eq:pregradphi_single} $N_c$ is the total number of configurations created.

\section{\label{sec:iter} Crystal structure prediction including ionic anharmonicity with machine-learning interatomic potentials}

\subsection{\label{sec:itreative}Iterative learning}

The approach proposed in this work focuses on training MLIPs that are both broadly applicable across the entire configuration space as well as accurate enough for further thermodynamic calculations at the anharmonic level of stable and metastable phases defined by a specific set of elements. To meet these requirements, we combine iterative MLIP training with an EA-based CSP, starting from the MatterSim foundation model \cite{yangMatterSimDeepLearning2024}. Using a robust foundation model as a starting point allows us to relax random crystals at high pressures robustly without an initial finetuning, bypassing the need to train MLIPs on random, unrelaxed structures with high energies, thereby largely reducing the size of training data. This feature distinguishes our approach from previous MLIP-assisted CSP methods~\cite{pickardEphemeralDataDerived2022,salzbrennerDevelopmentsFurtherApplications2023}. 

\begin{figure*}
    \includegraphics[width=0.60\textwidth]{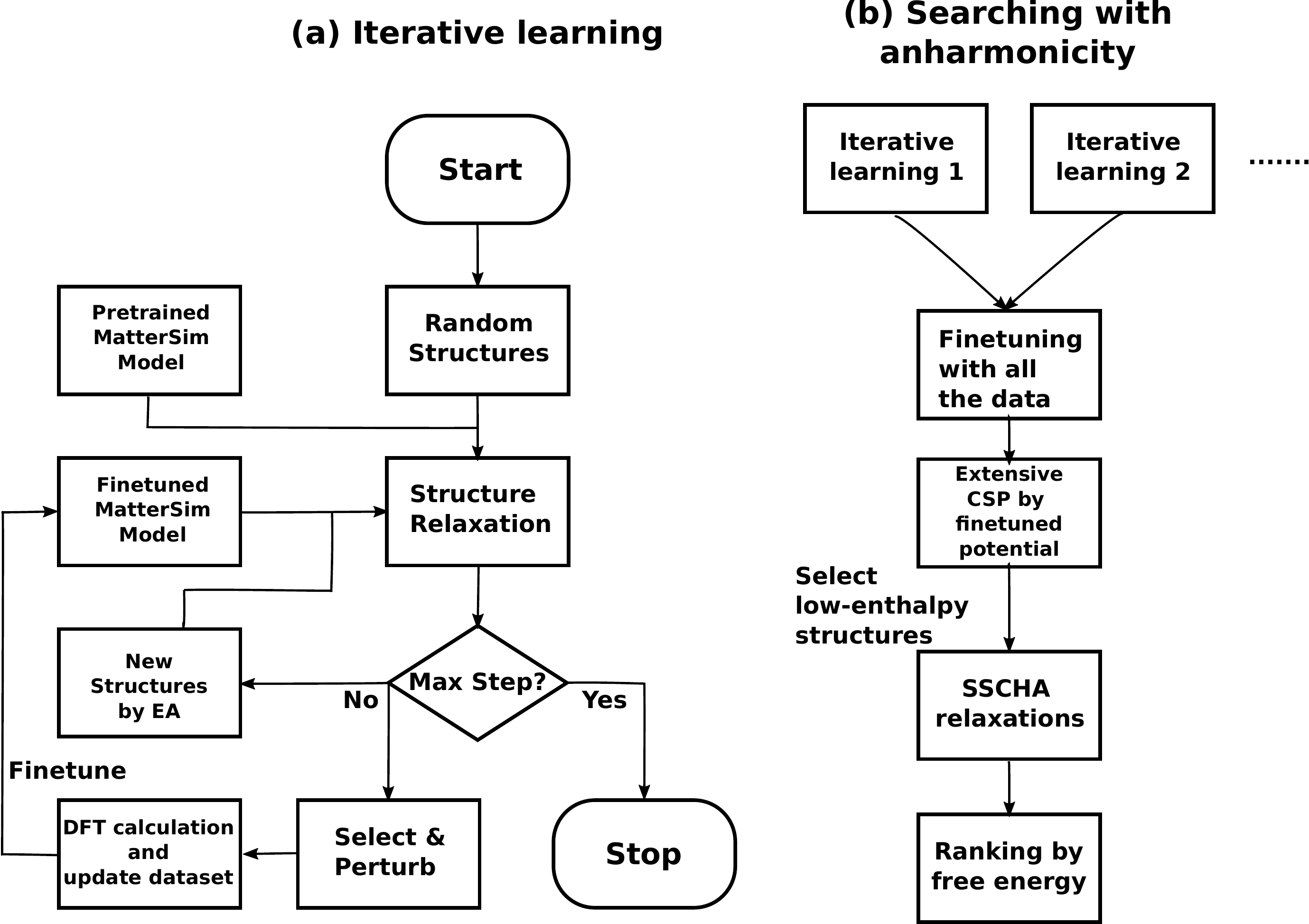}
    \caption{Iterative learning (a) and SSCHA-assisted searching (b) workflows in this work.}
    \label{fig:algo}
\end{figure*}

The iterative learning process (Fig. \ref{fig:algo}(a)) starts with randomly generated structures that are relaxed by the foundation Mattersim model in the first generation. Although MatterSim performs structure relaxations robustly under high pressures, it can still fail in predicting phase diagrams correctly, as shown in the following section. To address this issue, several of the obtained local minima are selected for a DFT sampling in order to finetune the model iteratively and improve its accuracy. Here we employ simple random selection while retaining duplicates in the population. This strategy increases the chances of selecting low-energy configurations, as they often appear multiple times due to evolutionary operators and basins of attraction on the PES during EA-based searching. Random distortions are applied to the atomic positions and cell parameters of the selected configurations to generate structures for single-point DFT calculations. The DFT energies, forces, and stresses obtained for them are used to update datasets. Because training large models on full historical data is computationally demanding, the updated dataset is formed by combining the new DFT data from the current generation with a random subset of the previous dataset. This combined dataset is then split into training, validation, and test sets, and MatterSim is finetuned on it. In parallel, new structures are generated by EA and relaxed using the finetuned MatterSim model, which are passed to the next generation. 
This search-and-learn cycle continues until a predefined maximum number of iterations is reached (Fig. \ref{fig:algo}(a)). 
If the target accuracy of the potential is not obtained, further cycles are performed.
This iterative learning is implemented (Fig. \ref{fig:algo}(a)) in MAGUS (Machine Learning and Graph Theory Assisted Universal Structure Searcher) code~\cite{xiaNovelSuperhardTungsten2018,gaoEnhancingCrystalStructure2021,wangMAGUSMachineLearning2023,hanEfficientCrystalStructure2025}. 

\subsection{\label{sec:sscha-search}SSCHA-assisted crystal structure searching}

During each iterative learning process, the target pressure is fixed. To extend the configuration space covered by generated datasets, multiple iterative learning processes could be carried out at different pressures, as shown in Fig. \ref{fig:algo}(b). The MatterSim model is then retrained on the combined dataset using the pretrained model as the starting point. The resulting well-trained MatterSim potential enables subsequent EA-based CSP searches with more generations for larger systems in a wide range of pressures, as well as accurate SSCHA relaxations of the generated structures in the CSP trajectory. 
In fact, once the potential is finetuned and acquires the desired accuracy, extensive CSP searches are performed at different pressures. 
In order to perform CSP at the anharmonic level, several low-enthalpy structures are selected and SSCHA relaxation are performed for them in order to obtain their free energies and rank their relative stabilities, as shown in Fig. \ref{fig:algo}(b). The SSCHA calculations can be performed at any temperature, starting from zero kelvin, effectively enabling finite temperature crystal structure predictions.

\section{Computational details}
\label{sec:detail}

\subsection{\label{sec:csp}Crystal structure prediction and iterative learning details}

We test the proposed approach on superconducting hydride H$_3$S\cite{duanPressureinducedMetallizationDense2014,drozdovConventionalSuperconductivity2032015,liDissociationProductsStructures2016,erreaQuantumHydrogenbondSymmetrization2016} which is well known as a highly anharmonic system. Initially, two independent iterative learning and searching processes are performed at 75 and 125 GPa starting from the MatterSim foundation model with nearly one million parameters(the model named ``MatterSim-v1.0.0-1M'' in MatterSim repository \cite{mattersimLink}). The number of atoms in the structures ranges from 4 to 32. Each search consists of 10 generations of iterative learning process (Fig. \ref{fig:algo}(a)), with 1000 structures relaxed per generation. Among them, 900 are randomly generated and 100 are produced by evolution operators, except for the first generation, in which all the structures are randomly generated. After relaxations, 10 structures are sampled randomly and each undergoes 10 different distortions: atomic positions shift by 0.05~\AA~in random directions while random strains from -5\% to 5\% are applied to cells. This step creates 100 distorted structures per generation for single-point DFT calculations. In addition to the newly computed DFT dataset, 100 configurations are randomly sampled from the previous training set, leading to 200 configurations used for finetuning. They are then divided into training, validation, and test sets in an 8:1:1 ratio. The MatterSim model is finetuned for 50 epochs with a learning rate of $2\times10^{-4}$ in each generation. 

Datasets from both searches are combined to finetune the final MatterSim foundation model as shown in Fig. \ref{fig:algo}(b). The final training, validation, and test datasets contain 1600, 200, and 200 configurations, respectively. The foundation model is firstly finetuned for 100 epochs with large force/stress loss ratios of 1, followed by further fine-tuning with force/stress loss ratios of 0.1. The learning rate is set to $1\times10^{-3}$. The resulting model is then employed for further calculations.

Finally, CSP under 50, 100 and 150 GPa is performed with the MAGUS code driven by the finetuned MatterSim model. For each pressure, more than 10000 configurations are sampled and relaxed. In order to determine the stable phases in a wider pressure range, low-enthalpy structures found are relaxed at more pressures between 50 and 200 GPa. Moreover, to incorporate anharmonicity, the 100 lowest-enthalpy structures at 100 GPa are selected and relaxed using SSCHA and the obtained finetuned potential at 0 K and a wide range of pressures.

\subsection{\label{sec:sscha_setting} Settings of the SSCHA calculations}

The SSCHA is used for evaluating thermodynamical stability and relaxing the crystal structures considering ionic quantum fluctuations and anharmonicity of various H$_3$S phases obtained by CSP. The SSCHA is also used to compute the free-energy Hessian matrices, which determine whether a structure is dynamically stable within the anharmonic free energy landscape~\cite{biancoSecondorderStructuralPhase2017}, of cubic H$_3$S both with the obtained MLIP through the procedure described in Sec. \ref{sec:iter} and  DFT. In the former task, variable-cell SSCHA relaxations were performed on $2\times2\times2$ supercells at 100 GPa for all the structures. The significant acceleration of the MLIP compared to DFT allows us to perform SSCHA calculations efficiently, even for structures that due to their low symmetry require a large number of configurations. We adopt a high-throughput approach for the SSCHA calculations by generating 1000 random configurations in each SSCHA population and limiting the SSCHA relaxation to a total of 30 populations, which guarantees a converged result. For free energy Hessian calculations of cubic $Im\bar{3}m$ H$_3$S, variable-cell SSCHA relaxations were first performed on $3\times3\times3$ supercells from 50 to 250 GPa with the MLIP. We also adopt a high-throughput approach here by making use of 1000 random configurations at each population and running in total 20 populations.  DFT-driven SSCHA relaxations were performed more carefully in order to save computational time. The SSCHA relaxations started from the auxiliary dynamical matrix computed by the MLIP with fixed cell parameters and $2\times2\times2$ supercells. The number of configurations is set to 100 or 200 per population, which are enough to converge results for this high-symmetry phase. Most DFT-SSCHA relaxations converged after 2 or 3 populations. Free-energy Hessian matrices and phonon frequencies \cite{biancoSecondorderStructuralPhase2017} were then obtained \emph{a posteriori} for both MLIP and DFT.

\subsection{\label{sec:dft}DFT settings}

The Perdew–Burke–Ernzerhof (PBE) functional \cite{perdewGeneralizedGradientApproximation1996} associated with pseudopotentials for H (1s$^2$) and S (3s$^2$3p$^4$) implemented in the VASP \cite{kresseEfficientIterativeSchemes1996,kresseEfficiencyAbinitioTotal1996,kresseUltrasoftPseudopotentialsProjector1999} code is used for DFT calculations. A plane-wave cutoff energy of 500 eV and a Brillouin zone integration grid spacing of 2$\pi \times 0.03$ \AA$^{-1}$ were used. Such setting is not of very high accuracy but it is practical and comparable with settings used for crystal structure searching of H$_3$S in previous works~\cite{duanPressureinducedMetallizationDense2014,liDissociationProductsStructures2016}.

\section{Results and Discussion}
\label{sec:res}

To evaluate the accuracy of the finetuned MatterSim potential to perform crystal structure predictions considering lattice anharmonicity as well as to estimate anharmonic vibrational frequencies, we use the well-studied H$_3$S system as a representative test case. According to enthalpy calculations in previous studies~\cite{duanPressureinducedMetallizationDense2014,erreaQuantumHydrogenbondSymmetrization2016,liDissociationProductsStructures2016}, the PBE functional predicts the following stability sequence for H$_3$S: $P1$ $\to$ $C2/c$ (2 GPa) $\to$ $R3m$ (112 GPa) $\to$ $I\bar{m}3m$ (175-180 GPa). The $I\bar{m}3m$ phase was experimentally identified as a high temperature superconductor at around 150 GPa~\cite{drozdovConventionalSuperconductivity2032015} but it presents imaginary modes below 173 GPa \cite{biancoHighpressurePhaseDiagram2018}, indicating that the structure is not a local minimum of the BO PES, in other words, that the structure is dynamically unstable in HA at the pressures where the highest superconducting critical temperature was observed. This discrepancy is resolved by SSCHA calculations that account for anharmonicity, revealing that the $I\bar{m}3m$ phase becomes dynamically stable above approximately 100 GPa~\cite{erreaQuantumHydrogenbondSymmetrization2016,biancoHighpressurePhaseDiagram2018}, much lower than the transition pressure predicted by static lattice calculations. Because H$_3$S is a challenging system for first-principles phonon calculations, it serves as an ideal benchmark for testing MatterSim’s accuracy in predicting vibrational properties and phase stability under high pressure at the anharmonic level.

\subsection{Test on the foundation model}

\begin{figure*}
    \includegraphics[width=0.50\textwidth]{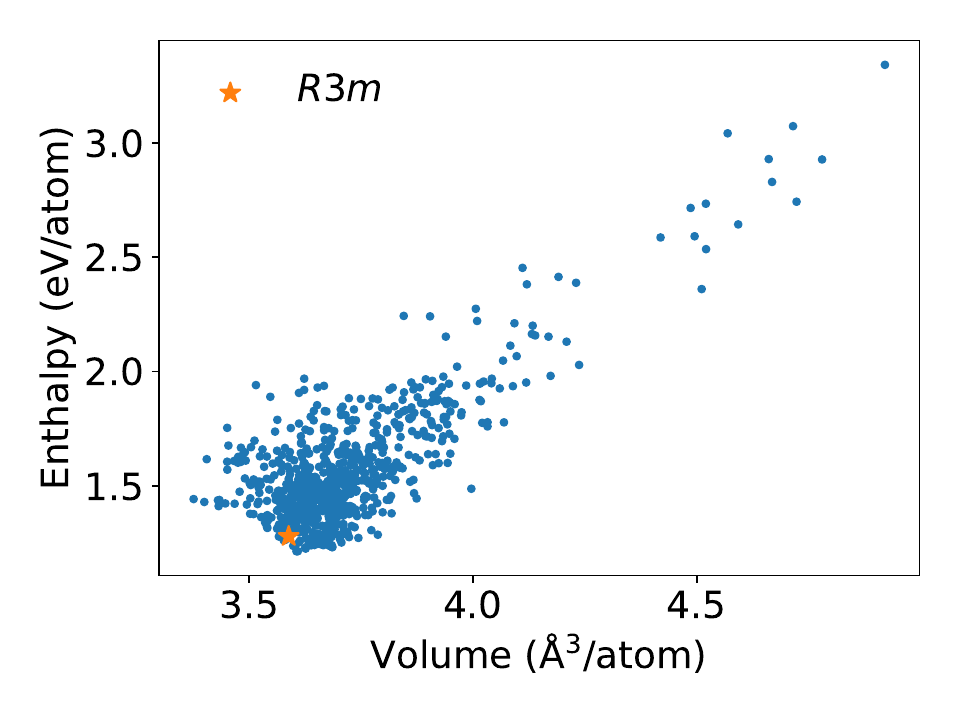}
    \caption{The enthalpy–volume plots for H$_3$S phases predicted by the pretrained MatterSim foundation model at 150 GPa.}
    \label{fig:ev}
\end{figure*}

The MatterSim foundation model enables robust relaxations from random initial structures, but fails to predict stable phases under high pressures. As shown in Table \ref{table:train}, the model exhibits energy root-mean-square errors (RMSEs) of approximately 80 meV/atom across training, validation, and test datasets. Such accuracy is too low for reliable stability determination given that the energy difference between competitive structures is of that order of magnitude~\cite{duanPressureinducedMetallizationDense2014}. For instance,
as shown in Fig. \ref{fig:ev}, the ground state $R3m$ phase predicted by DFT is 65 meV/atom above the lowest enthalpy crystal structure obtained by the foundation MatterSim model at 150 GPa and in the stability ranking it appears only at 60th place. This clearly highlights that foundation models by themselves are not enough for accurate thermodynamic stability calculations without further iterative training and finetuning.

\begin{table}
    \begin{ruledtabular}
        \caption{Root-mean-square errors of foundation and finetuned MatterSim models. Finetuned values are shown in bold. }
        \begin{tabular}{c c c c}%
           Dataset & Energy (meV/atom) & Force (meV/\text{\AA}) & Stress (GPa) \\
           \hline
           Training & 85.2/\textbf{6.1}& 398.1/\textbf{117.7}& 7.02/\textbf{0.48} \\
           Validation & 78.8/\textbf{6.6} & 388.0/\textbf{142.4} & 7.14/\textbf{1.14} \\
           Test & 85.3/\textbf{6.7} & 409.0/\textbf{136.0} & 6.88/\textbf{1.13} \\       
        \end{tabular}
        \label{table:train}
    \end{ruledtabular}
\end{table}

\subsection{\label{sec:train}Results obtained with the finetuned model}

\begin{figure*}
    \includegraphics[width=0.95\textwidth]{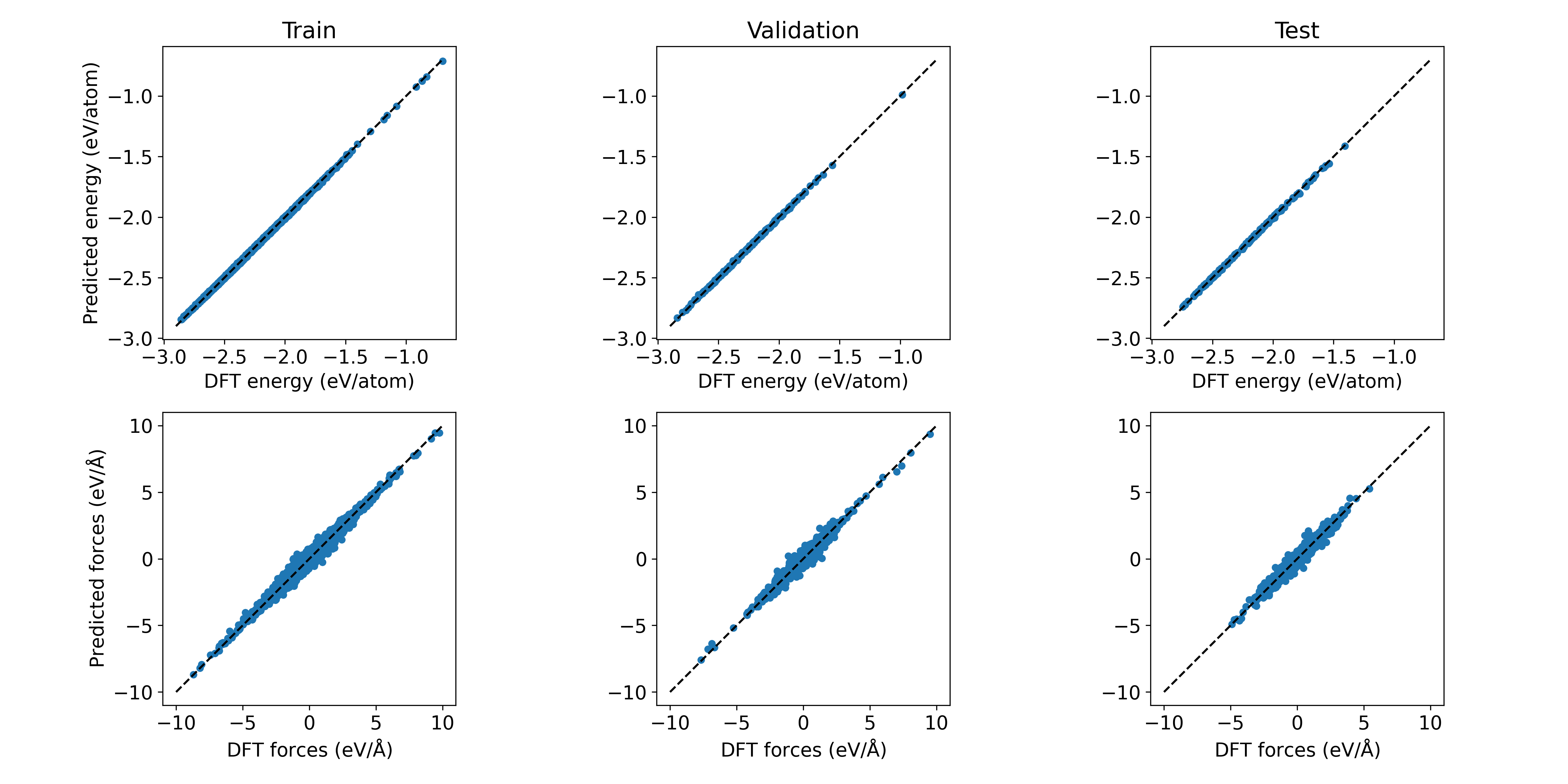}
    \caption{Parity plots of energy (upper) and forces (lower) comparing finetuned MatterSim predictions and DFT reference data for H$_3$S.}
    \label{fig:train}
\end{figure*}

The finetuning procedure sketched above significantly enhances the accuracy of the MatterSim potential for the H$_3$S system, reducing energy RMSEs to only $\sim$6 meV/atom, thus allowing to identify stable phases correctly and yielding an error that it is considerably lower than the impact of the ionic quantum fluctuations and anharmonicity on the total energy of the system~\cite{erreaQuantumHydrogenbondSymmetrization2016}. Energy, force, and stress comparisons between the finetuned MatterSim model and reference DFT calculations are presented in Table \ref{table:train} and Fig. \ref{fig:train}. While force RMSEs are larger than 100 meV/\text{\AA}, this performance remains acceptable considering the high diversity of the datasets, including molecular systems, metals, insulators, and various bonding patterns.

\subsection{\label{sec:vibs}Vibrational properties}

\begin{figure*}
    \includegraphics[width=0.5\textwidth]{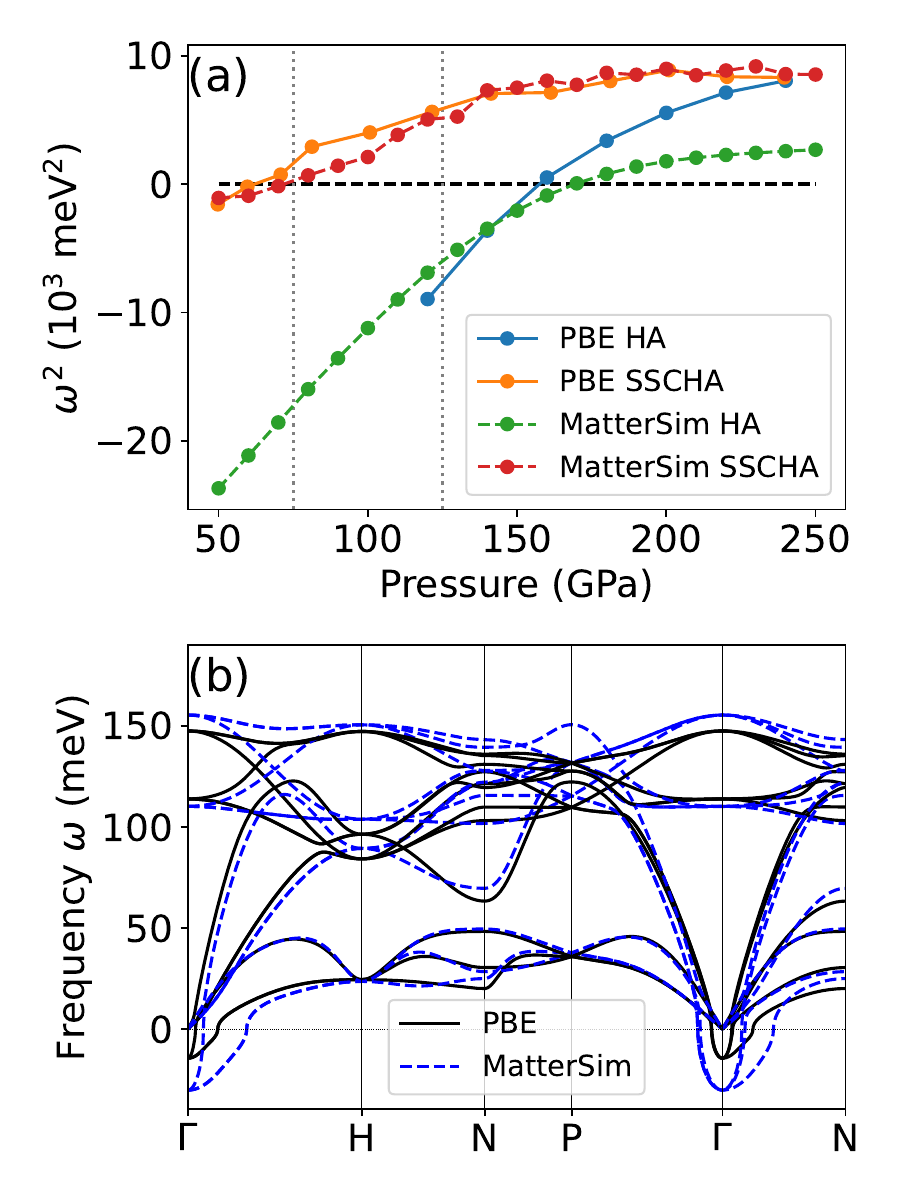}
    \caption{(a) Squared frequency of the $T_{1u}$ mode at $\Gamma$ point of H$_3$S $I\bar{m}3m$ phase, as a function of the pressure, predicted by the finetuned MatterSim and PBE with HA and SSCHA. The grey dashed lines represent pressures for generating datasets. (b) Anharmonic (at 0 K) phonon dispersions from free-energy Hessian matrices for H$_3$S computed by PBE and the finetuned MatterSim under 60 GPa. The imaginary mode at $\Gamma$ is the $T_{1u}$ mode.}
    \label{fig:vib}
\end{figure*}

Once the accuracy of the finetuned potential has been verified, we move forward and perform SSCHA calculations for the high-symmetry $I\bar{m}3m$ phase of H$_3$S to estimate its phonon frequencies as estimated from the anharmonic free energy Hessian~\cite{biancoSecondorderStructuralPhase2017,biancoHighpressurePhaseDiagram2018}. 
In this system, the soft optical mode $T_{1u}$ at $\Gamma$ within HA is related to the transition towards the $R3m$ phase. The squared frequencies of this mode as a function of pressure are computed using both the HA and the SSCHA. As shown in Fig. \ref{fig:vib}(a), the finetuned MatterSim predicts the critical pressure, where the frequency becomes zero, at the harmonic level in reasonable agreement with PBE-based  calculations. However, there are still discrepancies in the frequencies between PBE and MatterSim at higher pressures, with the MLIP tending to underestimate the frequency.

When the SSCHA is applied, the finetuned MatterSim and PBE results agree well across a broad pressure range. Although the training dataset was generated mainly at lower pressures (around 75 GPa and 125 GPa, indicated by dashed vertical lines in Fig. \ref{fig:vib}(a)), the finetuned potential accurately reproduces the anharmonic phonon frequencies over a much wider range, from 50 GPa to 240 GPa. The difference between the critical pressures predicted by PBE and the finetuned MatterSim is within 10 GPa.
Furthermore, as shown in Fig. \ref{fig:vib}(b), the potential can reproduce major features of renormalized phonon dispersions computed by PBE for this highly anharmonic system around the critical pressure. 
These results suggest that, interestingly, the finetuned MatterSim potential becomes more precise for SSCHA calculations than for standard harmonic calculations of phonon frequencies. 

\subsection{\label{sec:pd}Phase diagram with and without anharmonicity}

\begin{figure*}
    \includegraphics[width=0.5\textwidth]{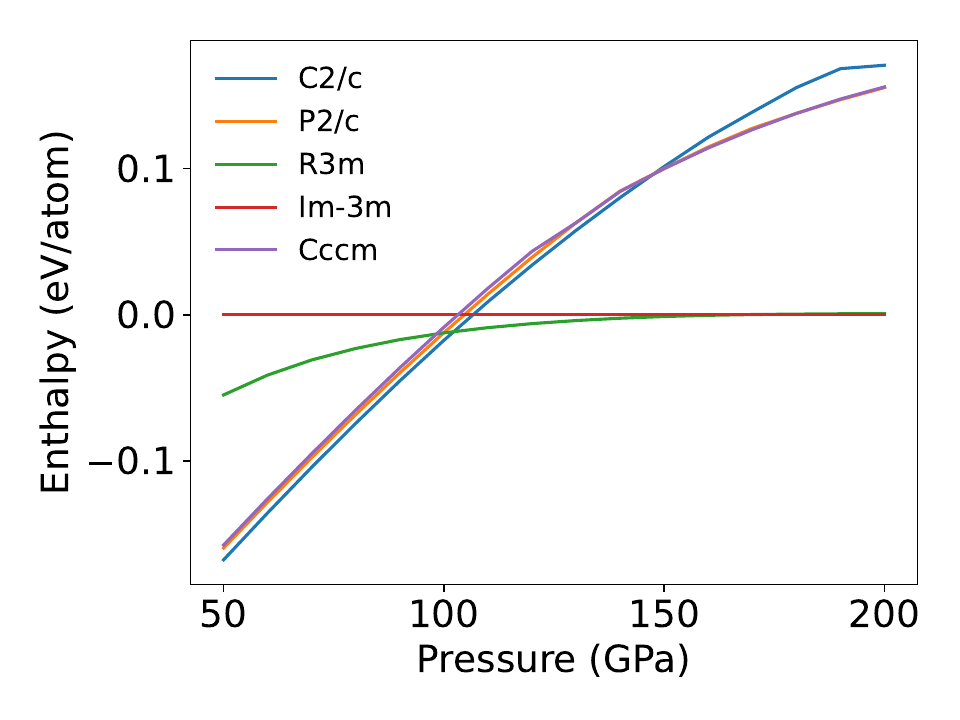}
    \caption{Relative enthalpy at the classical BO level without considering anharmonic effects of H$_3$S phases, as a function of the pressure, predicted by the finetuned MatterSim. The $I\bar{m}3m$ phase is set as the reference system.}
    \label{fig:enth}
\end{figure*}

\begin{figure*}
    \includegraphics[width=0.95\textwidth]{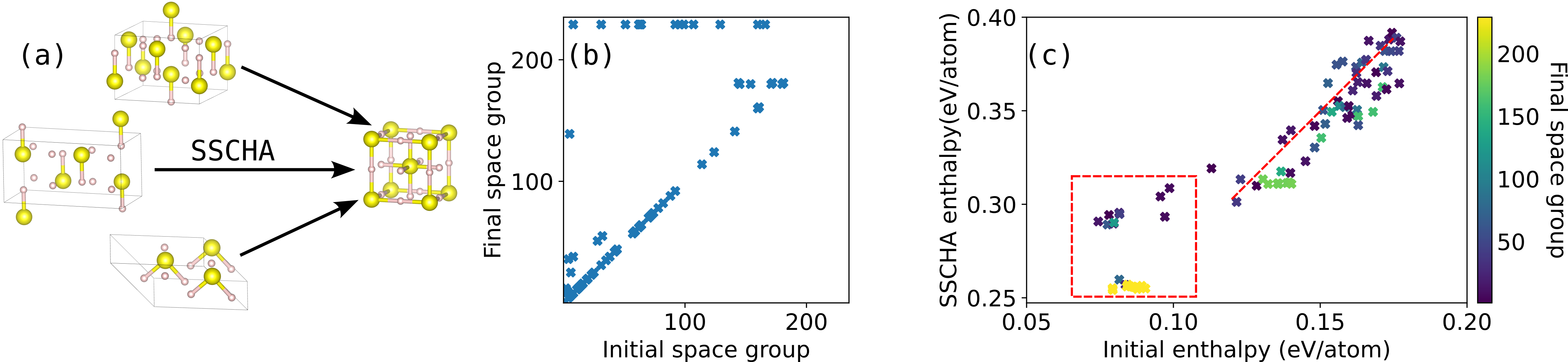}
    \caption{(a) Structural transformations from metastable H$_3$S phases to the cubic phase by SSCHA relaxations. (b) Symmetry analysis and (c) thermodynamic stability of H$_3$S phases before and after SSCHA relaxations at 100 GPa. In (c), the final space group of each relaxed structure is indicated by a colormap, with color corresponding to space group number. The red dashed box highlights a reordering of stability ranking in low-enthalpy region, while the red line indicates nearly linear correlation in high-enthalpy region.}
    \label{fig:scatter}
\end{figure*}

Next, we evaluate the accuracy of the obtained MLIP in phase diagram calculations with and without anharmonic effects. As shown in Fig. \ref{fig:enth}, the $C2/c$ phase is most stable below 100 GPa when the enthalpy is calculated from the BO PES without considering lattice anharmonicity~\cite{liDissociationProductsStructures2016}. This result is consistent with the transition pressure of 112 GPa predicted by PBE\cite{liDissociationProductsStructures2016}, indicating the predictive accuracy of the MLIP. Notably, the $C2/c$ phase is not in the training dataset, demonstrating the MLIP’s extrapolation capability to untrained structural regimes and pressures. 

To incorporate anharmonicity into the phase diagram, we first perform CSP at 100 GPa with the finetuned MatterSim. Then, we extract the 100 distinct structures with lowest enthalpies and perform fully variable-cell SSCHA relaxations at 0 K. This task is computationally prohibitive for DFT-based SSCHA and clearly benefits from the speed of MLIPs. As shown in Fig. \ref{fig:scatter}(a-b), 13 of these configurations transform into the high-symmetry $I\bar{m}3m$ phase after the SSCHA relaxation. To make sure that this phenomenon is not the artifact of the MLIP, we performed PBE-driven relaxations for these configurations. Most of them retained their original space group symmetry after subsequent PBE relaxations, demonstrating the reliability of the finetuned MatterSim model in predicting metastable phase configurations. Fig. \ref{fig:scatter}(c) confirms the cubic H$_3$S phase as the global minimum after SSCHA relaxations, extending previous findings on stabilization induced by quantum fluctuations\cite{erreaHighPressureHydrogenSulfide2015, erreaQuantumHydrogenbondSymmetrization2016} across a significantly broader free energy surface. Notably, Fig. \ref{fig:scatter}(c) also reveals that while SSCHA introduces corrections of around 200 meV/atom with respect to initial enthalpies, it significantly affects the stability ranking of configurations with low formation enthalpies (below 50 meV/atom, indicated by a red box). Beyond this region, the initial and SSCHA enthalpies exhibit a nearly linear correlation, highlighted by a red line in Fig. \ref{fig:scatter}(c). This indicates that anharmonic effects dominate only in low-enthalpy regimes, simplifying CSP with thermal and quantum fluctuations.

\subsection{\label{sec:reduce} Improved accuracy of the SSCHA calculations with the MLIP}

\begin{figure*}
    \includegraphics[width=0.8\textwidth]{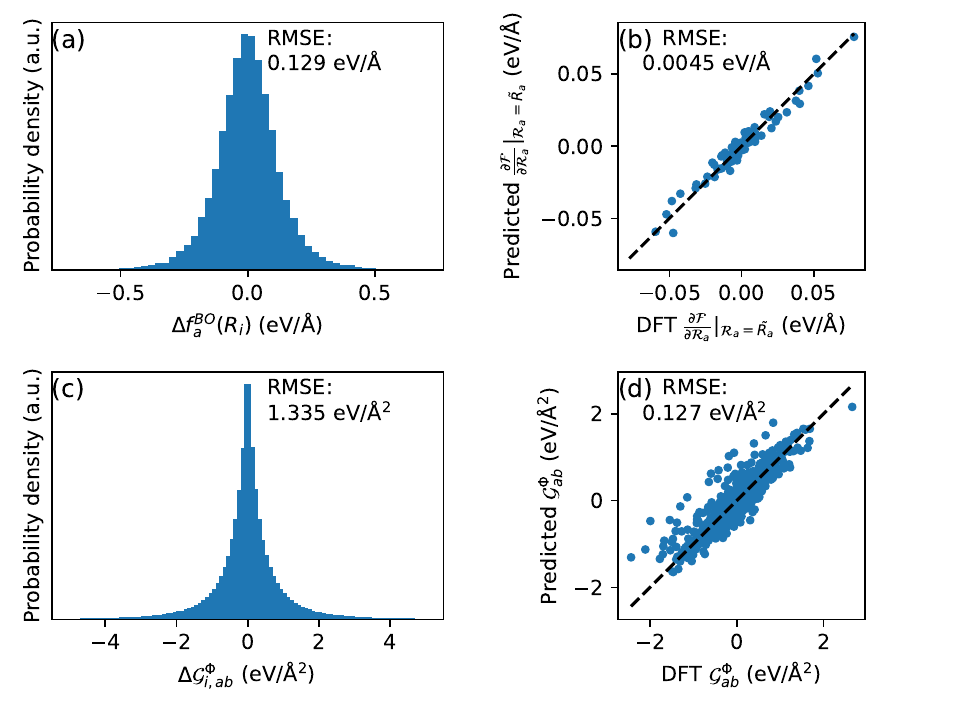}
    \caption{Distributions of gradient errors and parity plots comparing finetuned MatterSim predictions with DFT reference data for the $I\bar{m}3m$ phase of H$_3$S. (a) Distribution of $\{{\Delta \textbf{f}^{BO}(\bm{R}_i)}\}$. (b) Parity plot for $\frac{\partial \mathcal{F}}{\partial \boldsymbol{\mathcal{R}}} \Big|_{\bm{\mathcal{R}}=\tilde{\bm{R}}}.$ (c) Distribution of $\{\Delta \bm{\mathcal{G}}^{\bm{\Phi}}_{i}\}$. (d) Parity plot for $\bm{\mathcal{G}}^{\bm{\Phi}}$.}
    \label{fig:err}
\end{figure*}

\begin{figure*}
    \includegraphics[width=0.8\textwidth]{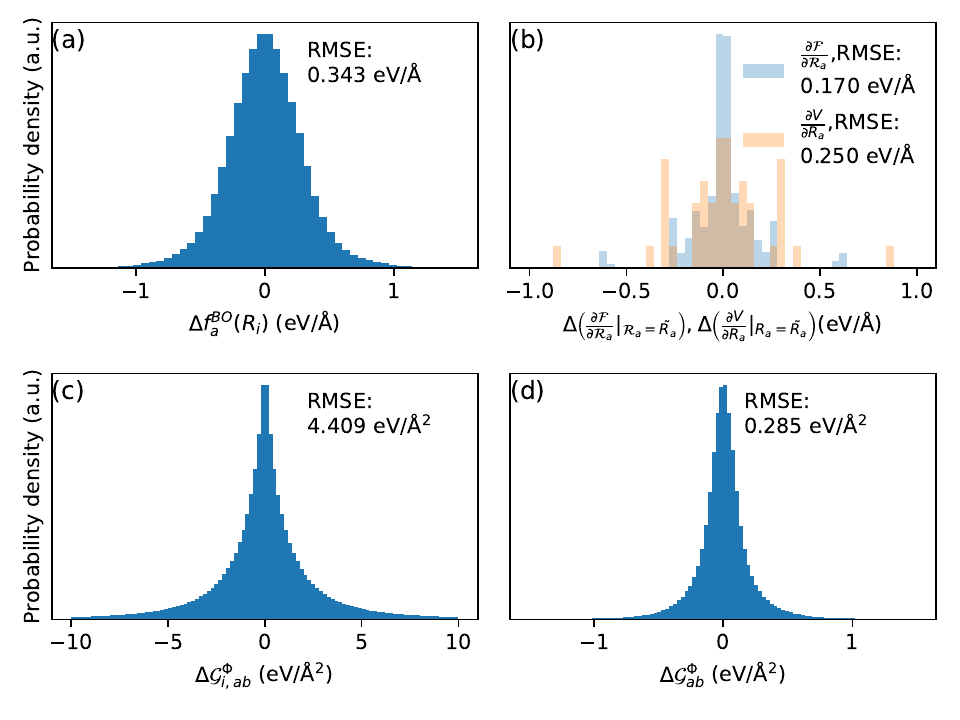}
    \caption{Distributions of gradient errors in the $C2/c$ phase of H$_3$S: (a) $\{{ \Delta \textbf{f}^{BO}(\bm{R}_i)}\}$, (b) $ \Delta \left( \frac{\partial \mathcal{F}}{\partial \bm{\mathcal{R}}} \Big|_{\bm{\mathcal{R}}=\tilde{\bm{R}}}  \right) $ and $ \Delta \left(\frac{\partial \mathcal{V}}{\partial\bm{R}} \Big|_{\bm{R}=\tilde{\bm{R}}}\right) $, (c) $\{{ \Delta \bm{\mathcal{G}}^{\bm{\Phi}}_{i}}\}$, (d) $ \Delta\bm{\mathcal{G}}^{\bm{\Phi}}$.}
    \label{fig:err_c2c}
\end{figure*}

As clearly shown above, the results for the vibrational properties obtained with the finetuned MLIP are closer to the DFT results at the SSCHA level than at the harmonic level. Also, the crystal structure search performed at the SSCHA level with the finetuned MatterSim potential is consistent with previous DFT+SSCHA results~\cite{erreaQuantumHydrogenbondSymmetrization2016}.   
All these observations suggest that, due to a certain cancellation of errors, SSCHA calculations tolerate moderate MLIP errors better than the standard BO PES calculations at the level of harmonic approximation, for both dynamical and thermodynamical properties.

In order to deepen on this idea, we first note that, as discussed in section~\ref{sec:sscha}, the performance of MLIP for SSCHA depends on its accuracy to evaluate the gradients of the SSCHA free energy. To compare the accuracy of the MLIP in SSCHA relaxation with that in a standard BO relaxation, we show below how the key quantities are computed.

Given centroids $\bm{\mathcal{R}}=\tilde{\bm{R}}$ and an auxiliary force constant matrix $\bm{\Phi}$, we generate an ensemble of configurations by sampling the associated SSCHA density matrix. For each configuration $i$, the BO forces $\textbf{f}^{\rm MLIP}(\bm{R}_i)$ and $\textbf{f}^{\rm DFT}(\bm{R}_i)$ are computed using DFT and the finetuned MLIP, respectively. The individual force errors are 
\begin{equation}
    \Delta \textbf{f}^{BO}(\bm{R}_i)  =  \textbf{f}^{\rm MLIP}(\bm{R}_i) - \textbf{f}^{\rm DFT}(\bm{R}_i),
    \label{eq:force_diff}
\end{equation}
and the full distribution of errors is obtained by collecting all components of $ \Delta \textbf{f}^{BO}(\bm{R}_i)$ across the ensemble. The RMSE of the forces is then given by 
\begin{equation}
{\rm RMSE}_f =\sqrt{\frac{1}{N}\sum_{i,a}\left|\Delta f^{\rm BO}_a(\mathbf{R}_i)\right|^{2}}, 
\label{eq:rmsef}
\end{equation}
where $a$ labels atomic force components and $N$ is the total number of components.

From the same ensemble, we can also compute the SSCHA free-energy gradients $\frac{\partial\mathcal{F}} {\partial\bm{\mathcal{R}}}\Big|_{\bm{\mathcal{R}}=\tilde{\bm{R}}}$ for the configuration $\tilde{\bm{R}}$ using both DFT and MLIP (Eq.~\ref{eq:gradR_simple}). The error in the free-energy gradient is defined as
\begin{equation}
    \Delta\left(\frac{\partial \mathcal{F}}{\partial \bm{\mathcal{R}}} \Big|_{\bm{\mathcal{R}}=\tilde{\bm{R}}} \right) = \frac{\partial \mathcal{F}^{\rm MLIP}}{\partial \bm{\mathcal{R}}}\Big|_{\bm{\mathcal{R}}=\tilde{\bm{R}}} - \frac{\partial \mathcal{F}^{\rm DFT}}{\partial \bm{\mathcal{R}}}\Big|_{\bm{\mathcal{R}}=\tilde{\bm{R}}}. 
    \label{eq:grad_diff}
\end{equation}
We also compute the BO forces $\frac{\partial V}{\partial \bm{R}}$ at the centroid positions $\tilde{\bm{R}}$ to compare the accuracy of MLIP for standard BO-based relaxation with that for SSCHA-based relaxation:
\begin{equation}
    \Delta \left(
    \frac{\partial V}{\partial\bm{R}} \Big|_{\bm{R}=\tilde{\bm{R}}}
    \right)
    =
    \frac{\partial V^{\rm MLIP}}
    {\partial\bm{R}} \Big|_{\bm{R}=\tilde{\bm{R}}}
    -
    \frac{\partial V^{\rm DFT}}
    {\partial\bm{R}} \Big|_{\bm{R}=\tilde{\bm{R}}}.
    \label{eq:centroid_force}
\end{equation}
Their distributions and RMSEs are evaluated in the same way as for the forces.

For the free-energy gradients with respect to the auxiliary force-constant matrix $\bm{\Phi}$, we calculate the per-configuration quantities and the averaged gradient
\begin{equation}
    \Delta \bm{\mathcal{G}}^{\bm{\Phi}}_i = \bm{\mathcal{G}}^{\bm{\Phi}, \rm MLIP}_i - \bm{\mathcal{G}}^{\bm{\Phi}, \rm DFT}_i,
    \Delta\bm{\mathcal{G}}^{\bm{\Phi}} = \bm{\mathcal{G}}^{\bm{\Phi}, \rm MLIP} - \bm{\mathcal{G}}^{\bm{\Phi}, \rm DFT},
    \label{eq:phi_diff}
\end{equation}
as defined in Eq.~\ref{eq:pregradphi_average} and \ref{eq:pregradphi_single} using forces computed by MLIP and DFT. Their error distributions and RMSEs are computed in the same way.

Based on the procedure described above, we can now evaluate how ensemble averaging affects the apparent accuracy of the MLIP. We first consider cubic $I\bar{m}3m$ H$_3$S at 60 GPa and 0 K. An ensemble of 500 configurations is generated using the converged auxiliary dynamical matrix from the SSCHA relaxation performed with PBE. Then, the forces of these configurations $\{{\textbf{f}^{BO}(\bm{R}_i)}\}$ are computed using both the finetuned MatterSim model and PBE. For this ensemble, the RMSE of the atomic forces (Eq.~\ref{eq:rmsef}) is 128 meV/\AA\ (Fig.~\ref{fig:err}(a)), consistent with those in Table.~\ref{table:train}.
However, when the same ensemble is used to evaluate the SSCHA free-energy gradient $\frac{\partial \mathcal{F}}{\partial \bm{\mathcal{R}}}$, the RMSE decreases drastically to only 4.5 meV/\AA\ (Fig.~\ref{fig:err}(b)). This large reduction comes from the nearly symmetric distribution of individual force errors (Fig.~\ref{fig:err}(a)), where over- and underestimations effectively cancel upon ensemble averaging. A similar compensation also occurs for free-energy gradients with respect to the auxiliary force-constant matrix $ \mathcal{G}^{\bm{\Phi}}_{ab}$, where averaged gradients show smaller errors compared to the individual contributions $\{ \mathcal{G}^{\bm{\Phi}}_{i,ab}\}$ (Fig.~\ref{fig:err}(c-d)). 

Due to the high symmetry of the $I\bar{m}3m$ phase, the gradients with respect to the centroids ideally vanish in practical SSCHA relaxations. The small residual values observed in Fig.~\ref{fig:err}(b) arise from finite sampling and the removal of symmetry constraints during averaging. To further assess the averaging effect in a lower-symmetry system, we performed the same analysis for the monoclinic $C2/c$ phase of H$_3$S at 100 GPa. Similar to the $I\bar{m}3m$ case, 500 configurations were generated, but here the auxiliary force-constant matrix and the configuration were optimized by SSCHA driven by the finetuned MatterSim model instead of PBE. Since the $C2/c$ phase was not included in the training data, the force errors in ${{\textbf{f}^{BO}(\bm{R}_i)}}$ are larger, reaching 343 meV/\AA\ (Fig.~\ref{fig:err_c2c}(a)). The ensemble-averaged gradient $\frac{\partial \mathcal{F}}{\partial \bm{\mathcal{R}}}$, however, shows a reduced error of 170 meV/\AA, corresponding to a 50\% decrease (Fig.~\ref{fig:err_c2c}(b)). Although this reduction is smaller than in the high-symmetry phase, a clear improvement is still achieved compared with the individual forces.

The RMSE of BO forces at centroids ($\frac{\partial \mathcal{V}}{\partial\bm{R}} \Big|_{\bm{R}=\tilde{\bm{R}}}$) is 250 meV/\AA~(Fig.~\ref{fig:err_c2c}(b)), which is higher than that of free-erngy gradient $\frac{\partial \mathcal{F}}{\partial \bm{\mathcal{R}}} \Big|_{\bm{\mathcal{R}}=\tilde{\bm{R}}}$ for the same structure (170 meV/\AA). This indicates that the MLIP reproduces free-energy gradients more accurately than potential-energy gradients, suggesting that SSCHA relaxations driven by MLIPs can be more reliable than standard relaxations. In some cases, MLIPs may even provide a more accurate description of the free-energy surface than the underlying BO-PES. Furthermore, although the errors in $\mathcal{G}^{\bm{\Phi}}_{ab}$ and $\{\mathcal{G}^{\bm{\Phi}}_{i,ab}\}$ are larger than those in the cubic phase, the relative error reduction is even more pronounced, as shown in Fig.~\ref{fig:err_c2c}(c–d). These results demonstrate that ensemble averaging in SSCHA provides robust error cancellation, even when the underlying MLIP force predictions are of limited accuracy.

\section{Conclusion}
\label{sec:end}

To enable CSP with anharmonic lattice dynamics, we proposed an iterative learning framework combining EA, an atomic foundation model, and SSCHA relaxations. MLIPs finetuned from foundation models significantly reduce required training data while enabling efficient SSCHA calculations to account for anharmonic effects. We validate this approach by successfully reproducing H$_3$S phase stabilities across a wide pressure range. 

Two key insights emerge for CSP with anharmonicity. Firstly, although SSCHA requires more static calculations than HA and standard relaxations, highly accurate SSCHA frequencies and free energies can be obtained by using a moderate-accuracy MLIP. This advantage arises from error compensation during ensemble averaging. Overestimations and underestimations cancel, reducing net errors in ensemble-averaged properties. This compensation remains even when the MLIP force predictions are relatively poor. Secondly, free energy corrections caused by anharmonic and temperature effects do not change the stability ranking dramatically even for highly anharmonic systems like H$_3$S. This insight suggests that SSCHA relaxations of low-enthalpy configurations are sufficient, avoiding exhaustive SSCHA calculations for all the configurations, thus enabling practical CSP with anharmonic lattice dynamics.

The proposed approach in this work paves the way for practical CSP with anharmonicity. While its performance at finite temperatures has not yet been tested, it offers a solid foundation for future extensions. Future work will focus on further reducing computational costs and enhancing the robustness of this approach for a broader range of materials.

\section{Acknowledgments}
This project is funded by the Spanish Ministry of Science and Innovation (Grant No. PID2022142861NA-I00) and the Department of Education, Universities and Research of the Eusko Jaurlaritza and the University of the Basque Country UPV/EHU (Grant No. IT1527-22).
 The authors acknowledge the financial support received from the IKUR Strategy under the collaboration agreement between Ikerbasque Foundation and Centro de Física de Materiales (CFM-MPC) on behalf of the Department of Science, Universities and Innovation of the Basque Government (HPCAI21: AI-CrysPred).
This project is also partially supported by the Extraordinary Grant of CSIC (Grant No. 2025ICT122).
Y.-W.F. and I.E. acknowledge PRACE for awarding access to the EuroHPC supercomputer LUMI located in CSC's data center in Kajaani, Finland through EuroHPC Joint Undertaking (EHPC-REG-2022R03-090 and EHPC-REG-2024R01-084). Technical and human support provided by DIPC Supercomputing Center is gratefully acknowledged.
The authors acknowledge enlightening discussions with the partners of the SuperC collaboration. The authors thank {\DJ}or{\dj}e Dangi{\'c} at UPV/EHU and Yu Han at Nanjing University for helpful discussion.

\bibliography{ref_H3S}

\end{document}